\documentclass[creativecommons,copyright]{eptcs}
\providecommand{\event}{GAM 2015} 

\usepackage{substr}

\usepackage{latexsym}

\usepackage{xcolor}
\usepackage{xspace}

\usepackage{framed}
\usepackage[T1]{fontenc}
\usepackage{graphicx}
\usepackage[publisher]{prov}

\usepackage{algpseudocode}
\usepackage{algorithmicx}
\usepackage{algorithm}

\usepackage{hyperref}

\usepackage{datagraphy}

\hypersetup{
     colorlinks   = true,
     citecolor    = blue,
     linkcolor=blue,
     urlcolor=blue}

\newcommand{\wdf}{\mathit{wdf}}
\newcommand{\wifb}{\mathit{wifb}}

\title{\provTitle{Aggregation by Provenance Types: A Technique for Summarising Provenance Graphs}\footnote{\provBanner{http://eprints.soton.ac.uk/364726/63/provenance.ttl}}}
\author{\provAuthor{Luc Moreau}{http://www.ecs.soton.ac.uk/\twidle lavm/}
\institute{Web and Internet Science\\ Electronics and Computer Science}
\institute{\provOrganization{University of Southampton}{http://www.soton.ac.uk/},
Southampton, UK}
\email{l.moreau@ecs.soton.ac.uk}
}
\def\titlerunning{Aggregation by Provenance Types}
\def\authorrunning{Luc Moreau}
\begin{document}
\maketitle
\provThis%
\provSpecialization{06EB3FFB-87EF-4A24-BE18-2D3817DC826A}
\newtheorem{req}{Requirement}

\newcommand{\APT}{\mbox{\sc apt}\xspace}

\newcommand{\ato}{{\sc ato}\xspace}
\newcommand{\col}{{\sc col}\xspace}
\newcommand{\lin}{{\sc lin}\xspace}
\newcommand{\pco}{{\sc pc1}\xspace}
\newcommand{\pon}{{\sc pon}\xspace}
\newcommand{\pri}{{\sc pri}\xspace}

\newcommand{\RDF}{{\sc rdf}\xspace}
\newcommand{\REST}{{\sc rest}\xspace}
\newcommand{\POST}{{\sc post}\xspace}

\newcommand{\etal}{{\em et al.}\xspace}

\newcommand{\Method}{\noindent{\bf Method}\xspace}
\newcommand{\Analysis}{\noindent{\bf Analysis}\xspace}

\newcommand{\RESTful}{{\sc rest}ful\xspace}
\newcommand{\API}{{\sc api}\xspace}
\newcommand{\URL}{{\sc url}\xspace}
\newcommand{\HTTP}{{\sc http}\xspace}
\newcommand{\HTML}{{\sc html}\xspace}
\newcommand{\PROV}{{\sc prov}\xspace}
\newcommand{\PROVDM}{{\sc prov-dm}\xspace}
\newcommand{\PROVN}{{\sc prov-n}\xspace}
\newcommand{\PROVJSON}{{\sc prov-json}\xspace}
\newcommand{\PROVXML}{{\sc prov-xml}\xspace}
\newcommand{\PROVCONSTRAINTS}{{\sc prov-constraints}\xspace}

\newcommand{\rdfproperty}[2]{\href{#2}{#1}}

\newcommand{\provbook}[1]{\rdfproperty{\tt provbook:#1}{http://www.provbook.org/#1}}
\newcommand{\bk}[1]{\rdfproperty{\tt bk:#1}{http://www.provbook.org/ns/\##1}}
\newcommand{\now}[1]{\rdfproperty{\tt now:#1}{http://www.provbook.org/nownews/#1}}
\newcommand{\nowis}[1]{\rdfproperty{\tt is:#1}{http://www.provbook.org/nownews/is\##1}}
\newcommand{\nowpeople}[1]{\rdfproperty{\tt nowpeople:#1}{http://www.provbook.org/nownews/people/#1}}
\newcommand{\othernews}[1]{\rdfproperty{\tt other:#1}{http://www.provbook.org/othernews/#1}}
\newcommand{\policy}[1]{\rdfproperty{\tt pol:#1}{http://www.provbook.org/policyorg/#1}}
\newcommand{\gov}[1]{\rdfproperty{\tt gov:#1}{http://www.provbook.org/gov/#1}}
\newcommand{\govftp}[1]{\rdfproperty{\tt govftp:#1}{ftp://ftp.bls.gov/pub/special.requests/oes/#1}}
\newcommand{\prov}[1]{\rdfproperty{\tt prov:#1}{http://www.w3.org/ns/prov\##1}}
\newcommand{\ann}[1]{\rdfproperty{\tt ann:#1}{http://provenance.ecs.soton.ac.uk/annotate/ns/\##1}}

\newcommand{\typeof}[1]{T[#1]}
\newcommand{\fail}{{\sf fail}}
\newcommand{\true}{{\sf true}}

\newcommand{\entity}{ent\xspace}
\newcommand{\agent}{ag\xspace}
\newcommand{\activity}{act\xspace}
\newcommand{\used}{use\xspace}
\newcommand{\actedOnBehalfOf}{del\xspace}
\newcommand{\wasAttributedTo}{attr\xspace}
\newcommand{\wasInformedBy}{comm\xspace}
\newcommand{\wasInfluencedBy}{infl\xspace}
\newcommand{\wasEndedBy}{end\xspace}
\newcommand{\wasStartedBy}{start\xspace}
\newcommand{\wasInvalidatedBy}{inv\xspace}
\newcommand{\wasGeneratedBy}{gen\xspace}
\newcommand{\wasDerivedFrom}{der\xspace}
\newcommand{\wasAssociatedWith}{assoc\xspace}
\newcommand{\specializationOf}{spec\xspace}
\newcommand{\alternateOf}{alt\xspace}
\newcommand{\hadMember}{mem\xspace}
\newcommand{\relation}{relation\xspace}

\newcommand{\rest}[1]{|\ #1}

\newcommand*\Let[2]{\State #1 $\gets$ #2}

\newtheorem{hypothesis}{Hypothesis} 
\newtheorem{dfn}{Definition} 
\newtheorem{lem}{Lemma}

\newcommand*\DNA{\textsc{dna}}

\newcommand{\up}[1]{\begin{sideways}$#1$\end{sideways}}

\begin{abstract}
As users become confronted with a deluge of provenance data, dedicated
techniques are required to make sense of this kind of information.  We
present {\em Aggregation by Provenance Types\/}, a provenance graph
analysis that is capable of generating provenance graph summaries. It
proceeds by converting provenance paths up to some length $k$ to
attributes, referred to as provenance types, and by grouping nodes
that have the same provenance types. The summary also includes numeric
values representing the frequency of nodes and edges in the original
graph. A quantitative evaluation and a complexity
analysis show that this technique is tractable; with small values
of $k$, it can produce useful summaries and can help detect
outliers. We illustrate how the generated summaries can further be
used for conformance checking and visualization.
\end{abstract}
%



\section{Introduction}\label{intro:section}

In the world of art, the notion of provenance is well understood.  A
piece of art sold in an auction is typically accompanied by a paper
trail, documenting the chain of ownership of this artifact, from its
creation by the artist to the auction. This documentation is referred
to as the provenance of the artifact.  Provenance allows experts to
ascertain the authenticity of the artifact, which in turn influences
its price. This paper is concerned with an
electronic representation of provenance for data, documents, and
in general things in the world.

Provenance is a record that describes how entities, activities, and agents have influenced a piece of data~\cite{Moreau:prov-dm:20130430}.  It can help users make trust judgements about data. \PROV is a set of W3C specifications~\cite{Groth:prov-overview:20130430}  aiming to facilitate the representation and exchange of provenance on the Web.  \PROV is a graph-based data model that is domain-agnostic and has been applied to a wide range of applications, including climate assessment\footnote{\url{http://nca2014.globalchange.gov/report}}, legal notices\footnote{\url{https://www.thegazette.co.uk/}}, crowd sourcing~\cite{Huynh:HCOMP13}, and disaster response~\cite{atomic:2012}. 

As \PROV gets  broader adoption, increasing numbers of applications\footnote{\url{https://sites.google.com/site/provbench/provbench-at-bigprov-13}.} continuously generate provenance, leaving users with a ``provenance  data deluge'' making it challenging for them to make sense of this information.  Users are confronted to questions such as Q1: ``What is the essence of the provenance presented to me?'', Q2: ``Is today's behaviour conformant to yesterday's provenance" or  Q3: ``Are there anomalies or outliers in an execution that deserve further investigation?". Such questions  become very quickly untractable as provenance traces grow, and their graphical representation fills entire screens.

Network analysis techniques can effectively process graphs; for instance, graph clustering seems a particularly relevant approach to address question Q1~\cite{Newman:2010:NI:1809753}. However, these techniques usually rely on statistical measures~\cite{Tian:2008}  and result in representations  where the meaning of provenance is lost.

Research in semi-structured databases~\cite{Buneman:1997,Nestorov:1998,Goldman:vldb1997} offers partial answers to our problem. Notions of graph schemas that can be extracted from graph data~\cite{Nestorov:1998,Goldman:vldb1997} can be used to summarise such data, offering an explanation about their structure, or helping formulate SPARQL queries~\cite{Campinas:Dexa2012,Jarrar:TKDE12}. Unfortunately, no approach is provenance-specific and capable of generating a summary that exposes the meaning of provenance relations. However, the concept of  conformance to a schema~\cite{Buneman:1997} is particularly relevant to addressing question Q2.

Visualization techniques use node size and line thickness for
quantitative comparisons and for conveying a sense of salience of
nodes and edges~\cite{Wattenberg:2006}. While our focus is not on
visualization, the idea of a weight to describe the
importance of an aspect is relevant.  This is a critical element that
can help identify outliers in large data sets~\cite{16937}, and  can
be the basis of a solution to Question Q3.

Meanwhile, there is a general trend of developing reusable and composable provenance graph transformations, for many purposes, including normalization~\cite{Cheney:prov-constraints:20130430}, obfuscation~\cite{Missier:TR2013}.  Our aim is to define a summarization functionality as a transformation, which can be used further, e.g., for visualization or conformance check. Specifically, we seek to generate a summary that has the following key characteristics:
\begin{enumerate}
\item A summary needs to preserve the meaning of provenance. Humans
  tend to aggregate historically distant entities, regarding them as
  ``similarly blurred in the past''.  But also a summary needs to
  distinguish entities that were obtained by distinct historical
  paths, since such paths may affect their trustworthiness.

\item A summary needs to contain numerical information about the
  frequency of elements of provenance graphs to enable outliers detection.
\end{enumerate}

Thus, we decided that ``provenance paths'', i.e. successions of provenance relations, should play a central role in our approach. For
instance,   {\em an entity derived from an entity attributed to an agent\/} is
a path of length 2. 
The intuition of our solution is to regard a provenance path of length
$k$ leading to a node as a ``provenance type'' for that node. Then, we construct a summary by aggregating all
nodes that have the same provenance types, and likewise for edges,
and by counting their occurrences.  The method,  named {\em
  Aggregation by Provenance Types\/}, $\APT(k$), is parametric in the
maximum path length. 
Specifically, the contributions of this paper are the following.
\begin{enumerate}
\item  A provenance graph analysis and transformation \APT that creates a summary of a provenance graph; this summary itself is built using the same provenance vocabulary, but should be seen as a provenance graph ``schema'' decorated with frequency information.
\item A complexity analysis of this algorithm, followed by a quantitative evaluations of the technique, showing its tractability, and by a discussion further showing its ability to expose salient properties of  provenance graphs.
\item An illustration of how \APT can be used in two applications such as visualization and conformance checking.
\end{enumerate}

The rest of this paper is structured as follows. First, we identify the
requirements that a summarisation technique is expected to meet
(Section \ref{requirements:section}). Then, we define the notion of
provenance type and the technique of ``aggregation by provenance
types'' (Section~\ref{apt:section}). It is followed by a complexity
analysis of the approach (Section~\ref{complexity:analysis}). We then
illustrate two different applications of \APT
(Section~\ref{applications:section}). This is followed by a
quantitative evaluation of the approach
(Section~\ref{evaluation:section}) and a discussion
(Section~\ref{discussion:section}).  Related work appears in
Section~\ref{related:section} and the paper finishes with concluding
remarks (Section~\ref{conclusion:section}).

\section{User Requirements for Summarisation}\label{requirements:section}

The three questions of Section~\ref{intro:section} form key user requirements that a provenance summarisation technique should satisfy.
The first requirement is inspired by Q1 of Section~\ref{intro:section} and some well-known requirements captured by the W3C Incubator group (see
\url{http://www.w3.org/2005/Incubator/prov/wiki/User_Requirements\#Use})
identifying the need to make provenance information understandable.
The following two requirements reflect directly Questions 2
and 3 of Section~\ref{intro:section}.

\begin{req}[Essence of Provenance]\label{essence:requirement}
A provenance summary should capture the essence of the provenance graph that it summarises.
\end{req}

\begin{req}[Conformance]\label{conformance:requirement}
It should be possible to decide whether a provenance graph is compatible, or conformant, with a provenance summary.
\end{req}

\begin{req}[Outliers]\label{outliers:requirement}
It should be possible to detect anomalies or outliers in a provenance summary.
\end{req}

We satisfy
Requirement~\ref{essence:requirement} by adopting the same language
for summaries as for provenance graphs.  
As we aggregate nodes and edges, we keep a count of the nodes and edges
that were collapsed; such numerical values can help a user or system
detect outliers. Finally, given that summaries are based on the same
language as provenance graphs, it becomes easy to check whether one is
compatible with the other; for instance, techniques such as simulations from
process algebra~\cite{Henzinger:95} can be applied to this problem \cite{Buneman:1997}.

Every day processes illustrate that we focus on more or less
recent past, according to how discriminating we want to be. For
instance, a selective graduate school may admit PhD applicants who
have a first-class degree from a reputable University. A more
selective graduate school will also require good transcripts from
secondary school, or extra-curricula activity related to the
subject. However, older information, such as performance in primary
school, is generally not used as discriminator. In manufacturing or food
production, the country of origin is deemed to be the place of last
substantial change. However, to avoid misleading food labels, the origin of
the primary ingredient should also be considered.  These two examples
show that individuals are distinguished according to their past ---
more or less recent.  Thus, the distance in the past can abstracted by
a parameter $k$, which we use in our aggregation method.

\section{Aggregation by Provenance Types}\label{apt:section}

The purpose of this section is to define {\em Aggregation by Provenance Types\/}, abbreviated by \APT.  To do so, we first introduce the notion of provenance type. 

A {\em provenance type\/} is defined as a category of things that have common characteristics from a provenance perspective. Provenance types are parameterised by an integer indicating the length of provenance paths used to characterise things. A level-0 provenance type is one of the core type predefined in \PROV: \prov{Entity}, \prov{Activity}, and \prov{Agent}, or any user-defined type. A level-$k+1$ provenance type is an expression describing the category of things one \PROV-relation away from things that have a level-$k$ provenance type.

For instance, in {\tt :a0 \prov{used} :e1}, if {\tt :e1} is assigned level-$k$ provenance type $\tau_k$, then {\tt :a0} is  assigned level-$k+1$ provenance type $used(\tau_k)$, meaning that {\tt :a0} belongs to the category of things that used something of type $\tau_k$. 
Likewise, in {\tt :e2 \prov{wasDerivedFrom} :e1}, if {\tt :e1} is assigned level-$k$ provenance type $\tau_k$, then {\tt :e2} is  assigned level-$k+1$ provenance type $\wdf(\tau_k)$, meaning that {\tt :e2} belongs to the category of things that were derived from something of type $\tau_k$. 

\begin{dfn}\label{provenance:type} A provenance type of level $k$, noted $\tau_k$, is defined as follows.
$$\begin{array}{rcl}
\tau_0&::=& Entity\ \ |\ \ Activity\ \ |\ \ Agent\ \ | \ \ \mbox{user-defined type}\\
\tau_{k+1}&::=&label(\tau_k) 
\end{array}
$$
where `$label$' is a \PROV property label  defined as follows.
\scriptsize$$\begin{array}{|l|l||l|l|}
\hline
\mbox{Label} & \mbox{\PROV property}&\mbox{Label} & \mbox{\PROV property}\\
\hline\hline
used &  \mbox{\prov{used}}&              wsb&  \mbox{\prov{wasStartedBy}}\\          
wgb& \mbox{\prov{wasGeneratedBy}}&      web&  \mbox{\prov{wasEndedBy}}\\            
\wdf&  \mbox{\prov{wasDerivedFrom}}&    \wifb&  \mbox{\prov{wasInformedBy}}\\        
waw&  \mbox{\prov{wasAssociatedWith}}&   mem&  \mbox{\prov{hadMember}}\\             
wat&  \mbox{\prov{wasAttributedTo}}&     spec&  \mbox{\prov{specializationOf}}\\     
aobo&  \mbox{\prov{actedOnBehalfOf}}&    alt&  \mbox{\prov{alternate}}\\             
wib&  \mbox{\prov{wasInvalidatedBy}}&&\\
\hline
\end{array}
$$
\end{dfn}

Note that Definition~\ref{provenance:type} lists forward properties of \PROV only, but could also consider \PROV inverse properties~\cite{Lebo:prov-o:20130430}. Provenance types with forward properties refer to what led to that node (i.e., the past of the node), whereas provenance types with inverse properties refer to what happened to that node (i.e., its future).

We introduce the properties \ann{pType0}, \ann{pType1},
\ldots that associate a resource with its provenance types of level 0,
1, \ldots, respectively.  A given resource may be the subject of
multiple provenance types properties. Figure~\ref{sparql:query}
displays a template of SPARQL query to compute provenance types.  

\begin{figure}[htb]
\begin{framed}
\scriptsize
\begin{verbatim}
PREFIX prov: <http://www.w3.org/ns/prov#>
PREFIX ann: <http://provenance.ecs.soton.ac.uk/annotate/ns/#>

CONSTRUCT {  ?y ?pType_kp1 ?provenanceType. }
WHERE {
   {
      ?y prov:wasDerivedFrom ?x.
      ?x ?pType_k ?t.
      BIND (CONCAT("wdf(",?t,")") AS ?provenanceType)
   }
   UNION
   {
      ?y prov:used ?x.
      ?x ?pType_k ?t.
      BIND (CONCAT("used(",?t,")") AS ?provenanceType)
   }
   // .... and similarly for other prov relations
}
\end{verbatim}
\end{framed}
\caption{Pattern of SPARQL Query to Compute Provenance Types of Level $k+1$. Provenance types are expressed as a string encoding of Definition~\ref{provenance:type}. }\label{sparql:query}
\end{figure}

\begin{figure}[ht]
\begin{framed}
\scriptsize
\begin{verbatim}
PREFIX prov: <http://www.w3.org/ns/prov#>
PREFIX ann: <http://provenance.ecs.soton.ac.uk/annotate/ns/#>

CONSTRUCT {  ?x ann:pType0 ?provenanceType. }
WHERE {
   {
     ?x a prov:Entity.
     ?x rdf:type ?b.
     BIND (CONCAT("",strafter(STR(?b),"#")) AS ?provenanceType)	
   }
   UNION
   // and similarly for prov:Agent and prov:Activity
}
\end{verbatim}
\end{framed}
\caption{Pattern of SPARQL Query to Compute Provenance Types of Level 0}\label{sparql:init}
\end{figure}

In
the query template, variables {\tt ?pType\_k} and {\tt ?pType\_kp1}
have to be instantiated to the properties reflecting the value $k$ of
interest: for instance, \ann{pType3} and \ann{pType4}, respectively,
for $k=3$.  A provenance type for level $k=0$ is computed by the
SPARQL query of Figure~\ref{sparql:init}, which assigns \ann{pType0}
to each resource of type \prov{Entity}, \prov{Activity}, or
\prov{Agent}. We assume that RDFS-reasoning has inferred \PROV core
types.

The graph transformation {\em Level-k Aggregation by Provenance Types\/}, written $\APT(k)$, constructs a provenance summary  by grouping all the nodes that have the same provenance types $\tau_i$ for any $i$ such that $0\leq i\leq k$, and then merging all edges, as specified in Definitions~\ref{node:aggregation} and~\ref{edge:aggregation}.
\begin{dfn}[Node Aggregation]\label{node:aggregation}
Given a \PROV document $D$, node aggregation for a value $k$ is a set of types  $T=\{t_0, t_1, ..., \}$ and a type assignment $f: Node\rightarrow T$, such that the following hold:
\begin{itemize}
\item If $f(x)=f(y)$ then $\phi(x,k)=\phi(y,k)$, for any node $x,y\in D$.

\item If $f(x)\not=f(y)$ then $\phi(x,k)\not=\phi(y,k)$, for any node $x,y\in D$.
\end{itemize}
where $\phi(x,k)=\{\tau_i | \tau_i\ \mbox{ is a provenance type of level $i$ for $x$, such that $0\leq i\leq k$}\}$.

Further, $\mbox{\it Nodes}: T \rightarrow N$,  the weight of each summary type, is defined as $\mbox{\it Nodes}(t)=|\ \{x\ |\ f(x)=t\}\ |$.
\end{dfn}

\begin{dfn}[Edge Aggregation]\label{edge:aggregation}
Let $D$ be a \PROV document. Let $T$ be a set of types $\{t_0, t_1, ..., \}$ and $f:
Node\rightarrow T$ be a type assignment obtained by node
aggregation for some $k$.
Let  $\psi(t_0,t_1,lab)$ be a set of edges $\{x \rightarrow^{lab} y\in D\ |\ f(x)=t_0  \mbox{ and } f(y)=t_1\}$.
Edge aggregation results in a weighted set of labeled edges, defined as follows:
$Edges: T\times T \times Label\rightarrow N$,
with
$Edges(t_0,t_1,lab)=|\psi(t_0,t_1,lab)| \mbox{ for any } t_0,$ $t_1,lab$.
\end{dfn}

\noindent From Definitions~\ref{node:aggregation} and~\ref{edge:aggregation}, we can now define \APT.

\begin{dfn}[Aggregation by Provenance Types]
Let $D$ be a \PROV document and level  $k\geq 0$ be a natural number.
Aggregation by Provenance Type, $\APT(k)(D)$ is a summary  $S$ such that:
\begin{itemize}
\item vertices of $S$ are the set $T=\{t_0, t_1, ..., \}$  obtained  by node aggregation for $k$, with type assignment $f$ mapping nodes of $D$ to $S$, and weight $\mbox{\it Nodes}$.
\item a weighted set of labeled edges $Edges$ obtained by edge aggregation for type assignment $f$.
\end{itemize}
\end{dfn}
As type labels ($t_0, \ldots$) are generated by \APT, summaries that only differ by
type labels are regarded as equivalent. Hence, we consider
summary equivalence up to type naming.

Provenance summaries are also graphs whose nodes and edges are
expressed according to \PROV.  
However, summaries in
general are not valid~\cite{Cheney:prov-constraints:20130430}
provenance graphs: indeed, collapsing nodes that have the same
provenance types may introduce cyclic derivations or specializations,
which are invalid~\cite{Cheney:prov-constraints:20130430}. For
instance, if a summary contains a triple {\tt :T
  \prov{wasDerivedFrom} :T}, it should be understood as: an entity of
type {\tt :T} was derived from another entity of type {\tt :T}.

Inspired by the notion of conformance of a graph to a schema~\cite{Buneman:1997}, we can
define a notion of conformance to a summary. (We assume that some nodes
have been identified as ``root'', from which all other nodes can be reached.)

\begin{dfn}[Conformance to a Summary]\label{summary:conformance}
A provenance graph instance $G$ conforms to a provenance graph summary $S$, in notation $G\preceq S$, if there exists a simulation from $G$ to $S$, i.e. a binary relation $\preceq$ from the nodes of $G$ to those of $S$ satisfying (1) the root nodes of $G$ and $S$ are in the relation $\preceq$, (2) whenever $u\preceq u'$ and $u\rightarrow^a v$ is an edge labeled $a$ in $G$, then there exists some edge between $u'$ and $v'$ for label $a$ in  $S$, i.e.  $Edges(u', v',a)>0$, such that $v\preceq v'$.
\end{dfn}

\begin{lem}
A provenance graph is conformant to any summary produced by $\APT(k)$ for any $k$.
\end{lem}
Proof is straightforward since the type assignment $f$ of
Definition~\ref{node:aggregation} provides a simulation from the graph
to its summary.

\section{Complexity Analysis}\label{complexity:analysis}

Following Definition~\ref{provenance:type}, the length of $\tau_{k+1}$
is bounded by the length of $\tau_k$ plus a constant. 
Thus, if $C_I$ is the maximum number of incoming edges per node, $N$ the number of nodes, and $c$ a constant, then the cost of computing all annotations for all nodes is:
$$cost(all\ \tau_{k})< N (C_I)^k  + c $$

\begin{figure*}[htb]
\scriptsize
\noindent\hspace{1.5cm}%
\begin{minipage}[b]{4.1cm}\scriptsize
\begin{verbatim}
:e1 a prov:Entity.
:e2 a prov:Entity.
:a a prov:Activity.
:ag a prov:Agent.
:e1 prov:wasGeneratedBy :a.
:e2 prov:wasGeneratedBy :a.
:a prov:used :e1.
:a prov:used :e2.
:e1 prov:wasAttributedTo :ag.
:e2 prov:wasDerivedFrom :e1.
\end{verbatim}
\end{minipage}\hspace{0.5cm}%
\provResource{https://eprints.soton.ac.uk/364726/14/example.pdf}
\includegraphics[width=0.14\hsize]{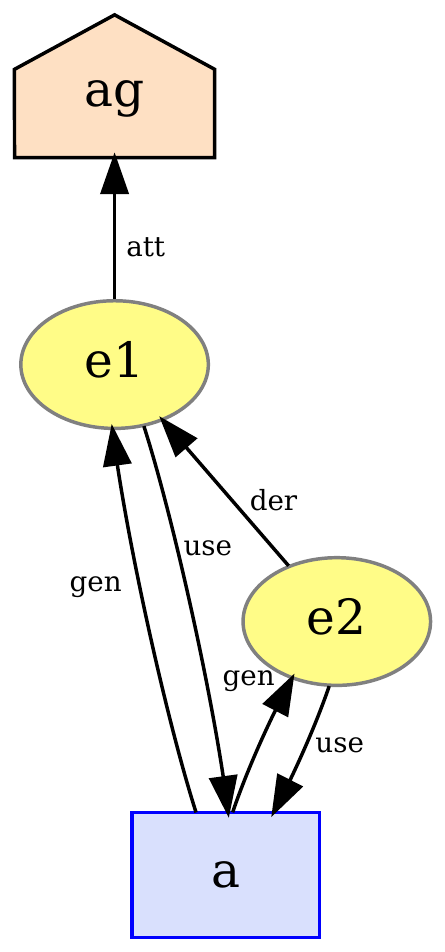}\hspace{0.5cm}%
\begin{minipage}[b]{5cm}\scriptsize
$$
\begin{array}[b]{l}
\mbox{provenance types for $a$}\\
\tau_0=Activity\\
\tau_1=\{used(Entity)\}\\
\tau_2=\{used(wat(Agent)),\\
\ \ \ \ \ \ \ \ used(\wdf(Entity))\}\\
\ \ \ \ \ \ \ \ used(wgb(Activity))\}\\
\ldots\\
\tau_4=\{\wdf(wgb(used(wat(Agent)))),\\
\ \ \ \ \ \ \ \ \wdf(wgb(used(\wdf(Entity)))),\\
\ \ \ \ \ \ \ \ \wdf(wgb(used(wgb(Activity)))),\\ 
\ \ \ \ \ \ \ \ wgb(used(\wdf(wat(Agent)))),\\
\ \ \ \ \ \ \ \ wgb(used(\wdf(wgb(Activity)))),\\
\ \ \ \ \ \ \ \ wgb(used(wgb(used(Entity))))\}\\
\ldots
\end{array}
$$
\end{minipage}
\caption{A Valid Provenance Graph with a Cycle and its Provenance Types}\label{cycle:example}
\end{figure*}

This means that computing provenance types is linear in the size of
the graph, but exponential in the analysis' level-$k$.  In the
presence of graphs with cycles (even valid
graphs~\cite{Cheney:prov-constraints:20130430}) the size of type
information may become exponentially large with
$k$. Figure~\ref{cycle:example} illustrates how provenance types grow
as $C_I$ is $2>1$, while the summary remains homomorphic to the original graph.

Note that this kind of complexity is typical of database graph schema techniques (see Section~\ref{related:section}). Furthermore, while \PROV allows for some valid graphs to be cyclic, this case is not so frequent.  In addition,
in Section~\ref{evaluation:section}, we show that \APT complexity is
practical. First, for a given level $k$, \APT is linear in the size of
the input graph. Second, summaries produced for $k=1$ and $k=2$ are
shown to be very usable. Third, we demonstrate that the size of the
summary saturates when $k$ is greater than the longest chain of
directed edges in the graph (see
Hypothesis~\ref{hypothesis:monotonic}).

In our proof of concept implementation, computation of provenance
types uses the SPARQL queries of Figures~\ref{sparql:query} and
\ref{sparql:init}. Generation of summaries and SVG representation rely
on ProvToolbox
(\url{http://lucmoreau.github.io/ProvToolbox/}). Ontology generation
uses the OWL-API.

\section{Applications of APT --- Illustration}\label{applications:section}

In this section, we provide two different applications of \APT:
interactive visualization and conformance check. We discuss them in turn.

\subsection{Visualization for Exploring Provenance}

Visualization tools for interactively exploring provenance could
leverage summaries.  While designing and implementing a summary-based
visualization tool is beyond the scope of this paper, one can outline
the appearance of a summary, and the modes of interaction that it
would support in a visualization tool.

Figure~\ref{scatter}
(right) illustrates how a a summary be visualized. There, the thickness of edges is determined by the relative
value of their weight as computed in the summary.  In a real
implementation, the size of nodes would also be indicative of their
respective frequencies.

As far as a visualization tool interface is concerned, we envisage two
adjacent windows: on the left-hand side, a summary representation, and
on the right-hand side, a graph instance.  Selection of nodes/edges in
the summary would automatically highlight corresponding nodes/edges in
instance.

\subsection{Conformance Check}

We revisit  Requirement~\ref{conformance:requirement}. A
provenance-enabled application generates a provenance trace, out of
which a summary is generated by \APT.  The next day, the application
continues to run, producing provenance again.  An expert user may
wonder whether the behaviour on the second say is conformant with the
application's behaviour on the first day.  Conformance
to a summary is defined in Definition~\ref{summary:conformance}.

To provide an \underline{\em illustration\/} of conformance checking,
we converted an \APT-summary into an OWL2 ontology (discarding all the frequency information).  In
Figure~\ref{owl2:definition}, we find two defined classes {\tt
  Execution\-Step\_5} and {\tt T\_9} from the example of Figure~\ref{scatter} (right).
The lines marked with $(\dagger)$ include basic types and definitions
of edges, as displayed in Figure~\ref{scatter}: for instance, {T\_9} is the source of a \prov{wasGeneratedBy} edge to either {\em ExecutionStep\_1\/} or {\em ExecutionStep\_5\/}.

Conformance checking is performed under the closed world assumption:
we consider here that we have complete knowledge. If a statement
cannot be found in a provenance graph, it is considered not to be
true.  Hence, the defined classes are extended with the clauses
$(\star)$ requiring the corresponding edges to be absent (maximum
cardinality is zero).  Similar clauses need to be generated for
instances, to express that there is no other possible instances that
can be asserted. Finally, all classes are defined to be mutually
disjoint.

Hence, an instance of activity {\em ExecutionStep\_5\/} is expected to have
used a {\em Building834.1\_2\/} and a {\em Route2324.0\_13\/}, and be
associated with {\em User\_6\/}.  The class {\em ExecutionStep\_5\/}
is distinct from {\em ExecutionStep514\_12\/} which has no usage, but
a similar association with {\em User\_6\/}.

\begin{figure*}[htb]
{\scriptsize
\begin{eqnarray*}
{ExecutionStep\_5}
&\ensuremath{\equiv}&\ensuremath{\exists}~used~Building834.1\_2 \ \ (\dagger)\\
&&\ensuremath{\sqcap}~\ensuremath{\exists}~used~Route2324.0\_13~\ \ (\dagger)\\
&&\ensuremath{\sqcap}~\ensuremath{\exists}~wasAssociatedWith~User\_6 \ \ (\dagger)\\
&&\ensuremath{\sqcap}~\ensuremath{\leq}~0~used\ensuremath{\lnot}(Building834.1\_2~\ensuremath{\sqcup}~Route2324.0\_13) \ \ (\star)\\
&&\ensuremath{\sqcap}~\ensuremath{\leq}~0~wasAssociatedWith~\ensuremath{\lnot}User\_6\ \ (\star)
\\
\\
T\_9&\ensuremath{\equiv}&~Vote\ \ (\dagger)\\
&&\ensuremath{\sqcap}~\ensuremath{\exists}~wasDerivedFrom~Building834.1\_2~\ \ (\dagger)\\
&&\ensuremath{\sqcap}~\ensuremath{\exists}~wasGeneratedBy~(ExecutionStep\_1~\ensuremath{\sqcup}~ExecutionStep\_5)\ \ (\dagger)\\
&&\ensuremath{\sqcap}~\ensuremath{\leq}~0~wasDerivedFrom~\ensuremath{\lnot}(Building834.1\_2)\ \ (\star)\\
&&\ensuremath{\sqcap}~\ensuremath{\leq}~0~wasGeneratedBy~\ensuremath{\lnot}(ExecutionStep\_1~\ensuremath{\sqcup}~ExecutionStep\_5)\ \ (\star)
\end{eqnarray*}
}
\caption{OWL2 Definition for ${ExecutionStep\_5}$ and $T_9$ from Figure~\ref{scatter} (right)}\label{owl2:definition}
\end{figure*}

\section{Evaluation}\label{evaluation:section}

In this section, we develop several hypotheses that we
validate by applying \APT to a set of provenance graphs and examining
the results. 
For the purpose of evaluation, we consider  provenance graphs generated by the following applications. 
\begin{enumerate}
\item  Atomic Orchid (\ato)~\datacite{AtomicOrchid} is a real-time location-based serious game to explore coordination and agile teaming 
 in disaster response scenarios~\cite{atomic:2012}.  Provenance in AtomicOrchid includes location and activities of participants, and orders issued by the headquarter.
\item CollabMap (\col)~\datacite{CollabMap} is an application that crowd-sources evacuation routes in a geographical area (with a view of simulating evacuations under various conditions)~\cite{Huynh:HCOMP13}. CollabMap provenance describes how all artifacts, i.e., building, routes, route sets, and votes, have been created.
\item Patina of Notes (\pon)~\datacite{Patina} is an application for collecting notes about archaeological artifacts, with a view to build, possibly multiple, interpretations of these artifacts~\cite{Jewell:IPAW2012}. The provenance includes the notes, their structures, and how they evolve over time.
\item The Provenance Challenge 1 (\pco)~\datacite{PC1} workflow is  representative of FMRI applications building brain atlases. It was the basis of the provenance challenge series and early provenance inter-operability efforts~\cite{Editorial:Challenge06}. 
\item The PROV Primer (\pri)~\datacite{Primer} describes the activities around the editing of a document~\cite{Gil:prov-primer:20130430}.
\item Linear Derivation (\lin)~\datacite{Linear} is a synthetic provenance graph exhibiting a
 linear sequence of successive derivations.
\end{enumerate}

The \APT graph transformation outputs a summary. Our first
investigation, captured by Hypothesis~\ref{compress:hypo}, is
concerned with the size of \APT's output. Intuitively, in the worst
case, when no node of the original graph can be aggregated by \APT,
the output has the same size as the input. In the best case, all nodes
can be merged in a single one; compression is then maximum.
\begin{hypothesis}[Compression]\label{compress:hypo}
Given a provenance graph $G$, \APT results in a summary whose number
of nodes is smaller than or equal to the number of nodes in the
original graph $G$.
\end{hypothesis}

\begin{figure}[htb]
\begin{center}
\provResource{https://eprints.soton.ac.uk/364726/7/compression-2.pdf}
\provFromData{https://eprints.soton.ac.uk/364726/8/ato.ttl}
\provFromData{https://eprints.soton.ac.uk/364726/9/col.ttl}
\provFromData{https://eprints.soton.ac.uk/364726/10/pon.ttl}
\provFromData{https://eprints.soton.ac.uk/364726/11/pc1.ttl}
\provFromData{https://eprints.soton.ac.uk/364726/12/pri.ttl}
\provFromData{https://eprints.soton.ac.uk/364726/13/lin.ttl}
\includegraphics[width=0.6\hsize]{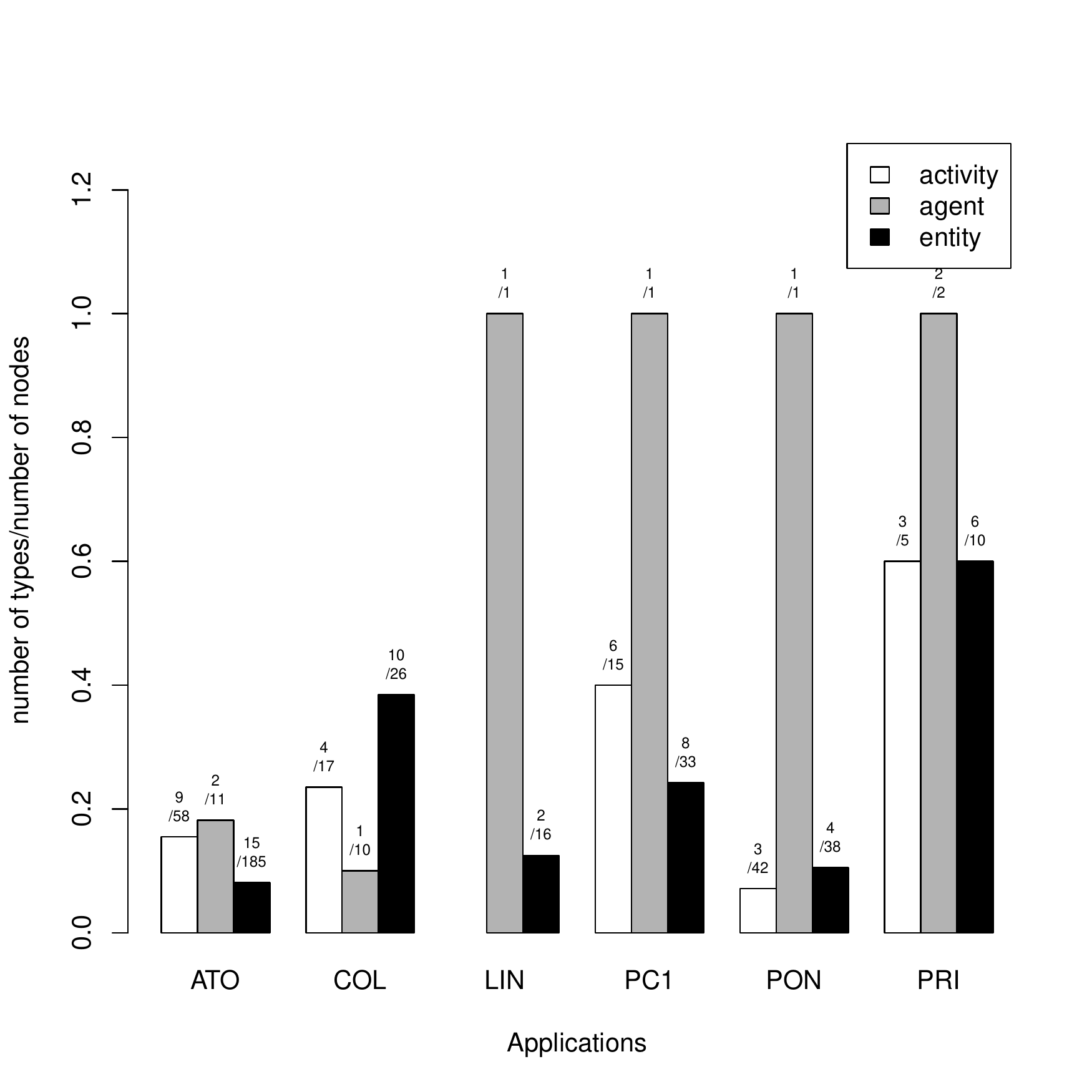}
\end{center}
\caption{Ratio ``number of output types : number of input nodes'' for $\APT(2)$. The smaller the more compressed.  Above each bar, we find the ratio $x/y$, where $x$ denotes the absolute number of  output types, whereas $y$ denotes the absolute number of input nodes. Plot can also be found \href{https://eprints.soton.ac.uk/364726/7/compression-2.pdf}{online}.}\label{compression-2}
\end{figure}

\Method We apply \APT to the selected set of provenance
graphs, and compare the number of types produced by \APT with the
number of input nodes, for types entity, activity, and agent.  To be
able to compare the relative performance of \APT on the different
graphs and different types, we plot the ratio ``number of types :
number of input nodes''. Figure~\ref{compression-2} plots such a ratio
for $\APT(2)$.    

\smallskip
\Analysis We see a significant reduction in size (compression ratio
between 3 and 10) for applications \ato, \col, \lin, \pco, \pon,
except for agents (which happen to be in small number for some graphs
and cannot be aggregated further).  The exception is
\pri, which describes a fairly unstructured series of human
activities related to editing a document: very few nodes can be
aggregated together, resulting in a small compression rate.$\Box$

\bigskip

Figure~\ref{compression-2} plots one point in the space of possible
outputs of \APT. As \APT level increases, the analysis is better able to
discriminate nodes that have a different history: thus, the
number of types also increases, until it saturates, as expressed by
Hypothesis~\ref{hypothesis:monotonic}. This hypothesis relies on a
graph metrics, Maximum Finite Distance, MFD\footnote{Specifically, we
  define MFD as the maximum of MFD entity to entity, MFD activity to
  entity, and MFD agent to entity.}~\cite{Ebden:IPAW2012}, which is a
provenance-specific variant of graph diameter.

\begin{hypothesis}[Monotonic]\label{hypothesis:monotonic}
The number of output types of $\APT(k)$ is a monotonically increasing
function of $k$ that plateaus once $k$ reaches the graph's Maximum
Finite Distance (MFD).
\end{hypothesis}

\begin{figure}[htb]
\begin{center}
\provResource{https://eprints.soton.ac.uk/364726/15/all-compression.pdf}
\includegraphics[width=0.6\hsize]{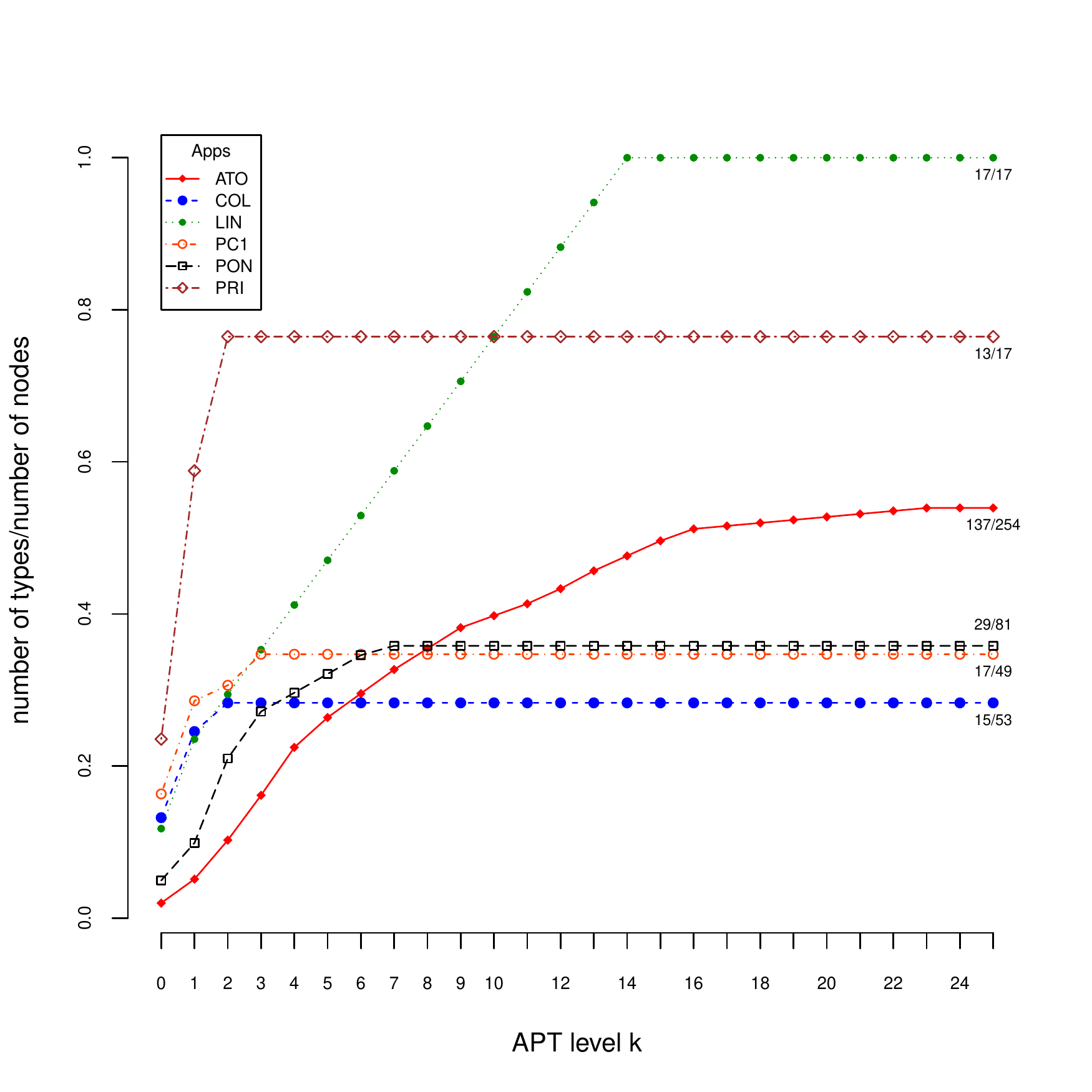}
\begin{minipage}[b]{4cm}
\begin{tabular}[b]{|l|c|c|}
\hline
app & plateau for $k$ & MFD~\cite{Ebden:IPAW2012}\\
\hline\hline
\ato & 24 & 24\\ 
\col & 2 & 4\\
\lin & 14 & 15 \\
\pco & 4 & 6\\
\pon &7& 8\\
\pri &2& 4\\
\hline
\end{tabular}
\vspace*{2cm}
\end{minipage}
\end{center}
\caption{Ratio ``number of output types : number of input nodes'' for
  $\APT(k)$, for increasing values of $k$. We also express the ratio $x/y$ when \APT plateaus (where $x$ denotes the number of  output types, whereas $y$ denotes the number of input nodes). Plot can also be found \href{https://eprints.soton.ac.uk/364726/15/all-compression.pdf}{online}.}\label{compression:function:of:n}
\end{figure}

\Method We apply \APT to the selected set of provenance
graphs, and compare the total number of types produced by $\APT(k)$ with the
total number of input nodes. Figure~\ref{compression:function:of:n} plots the ratio ``number of types :
number of input nodes'', for increasing values of $k$.

\smallskip
\Analysis For each application, $\APT(k)$ reaches a plateau for a
given value of $k$. The table in Figure~\ref{compression:function:of:n} records this value as well as
the MFD measure of the corresponding graph, computed according to~\cite{Ebden:IPAW2012}. We see that the plateau is
reached for a value of $k$ smaller than the graph's MFD.  Intuitively,
this can be explained by the fact that no distinct provenance type can be
propagated on chains longer than the Maximum Finite Distance.

In some cases, such as \lin, the types identified by \APT have a
direct connection with the nodes of the input graph.  In such
circumstances, with a value of $k$ that is large enough, \APT is
capable of distinguishing all nodes, and therefore does not aggregate
any. In other cases, such as \pco, no level of \APT is able to
distinguish some input nodes. Indeed, the workflow operates over a
collection of images (3 in this specific instance). Given that their
treatment is uniform, no \APT analysis can distinguish them, unless
they are explicitly assigned some distinct base types.$\Box$

\bigskip

The reason why \APT is capable of compressing the input graph (see
Hypothesis~\ref{compress:hypo}) is that \APT can
represent repeated graph patterns in a more compact manner in the
summary.  Hypothesis~\ref{hypo:repeated:patterns} relates numeric
values in the summary to the repeat of patterns in the original graph.

\begin{hypothesis}[Repeated Patterns]\label{hypo:repeated:patterns}
Types produced by $\APT(k)$ with occurrences $>k$ are likely to be
members of graph patterns repeated more than $k$ times.
\end{hypothesis}

\Method After applying \APT to a provenance graph, we plot the number
of nodes for each type identified by \APT.  The types
with number of nodes $>k$ are candidate nodes of repeated patterns.

\begin{figure}[htb]
\begin{center}
\provResource{https://eprints.soton.ac.uk/364726/16/scatter.pdf}%
\provFromData{https://eprints.soton.ac.uk/364726/9/col.ttl}%
\includegraphics[width=0.47\hsize]{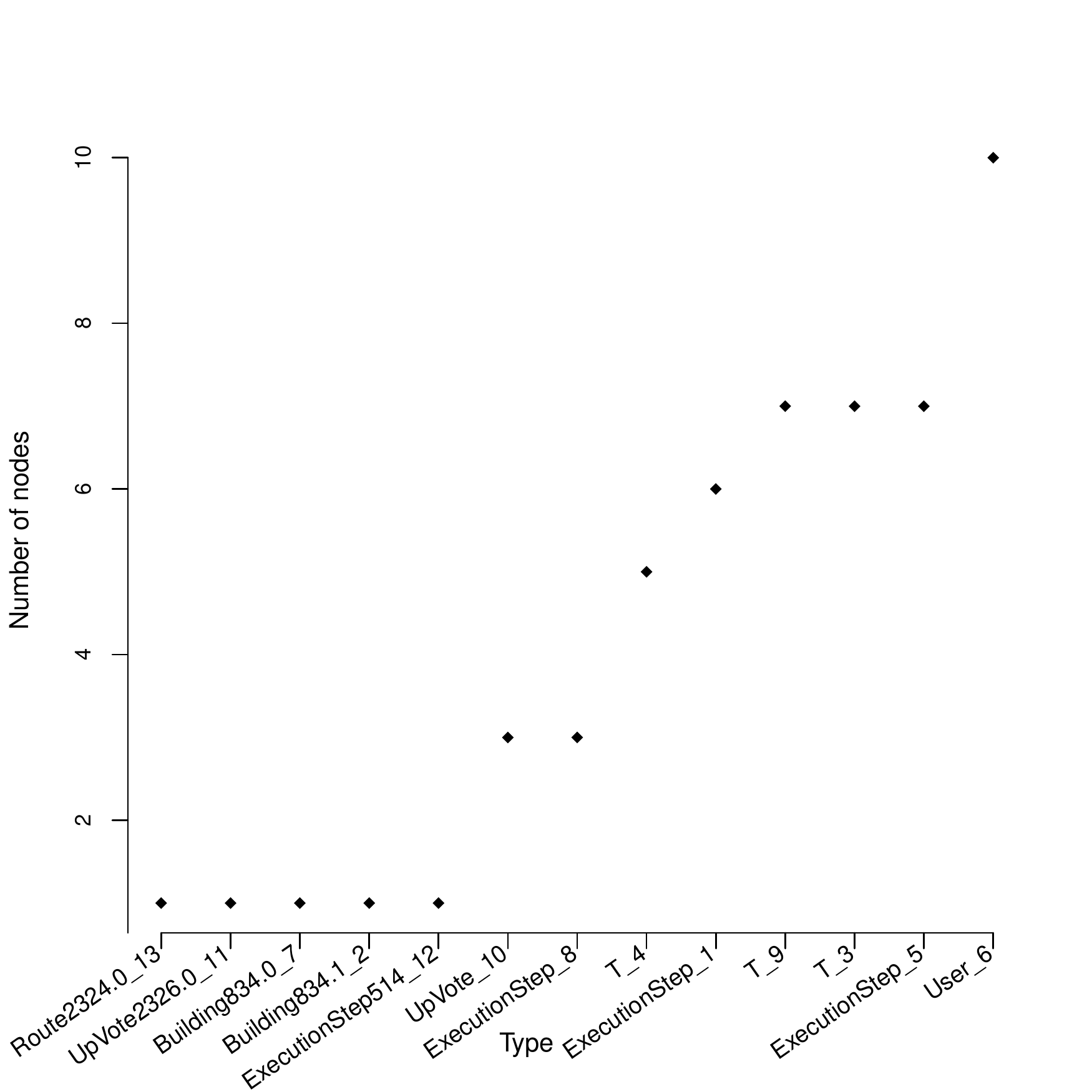}%
\provResource{https://eprints.soton.ac.uk/364726/17/collabmap-visual-1.pdf}%
\provFromData{https://eprints.soton.ac.uk/364726/9/col.ttl}%
\includegraphics[width=0.5\hsize]{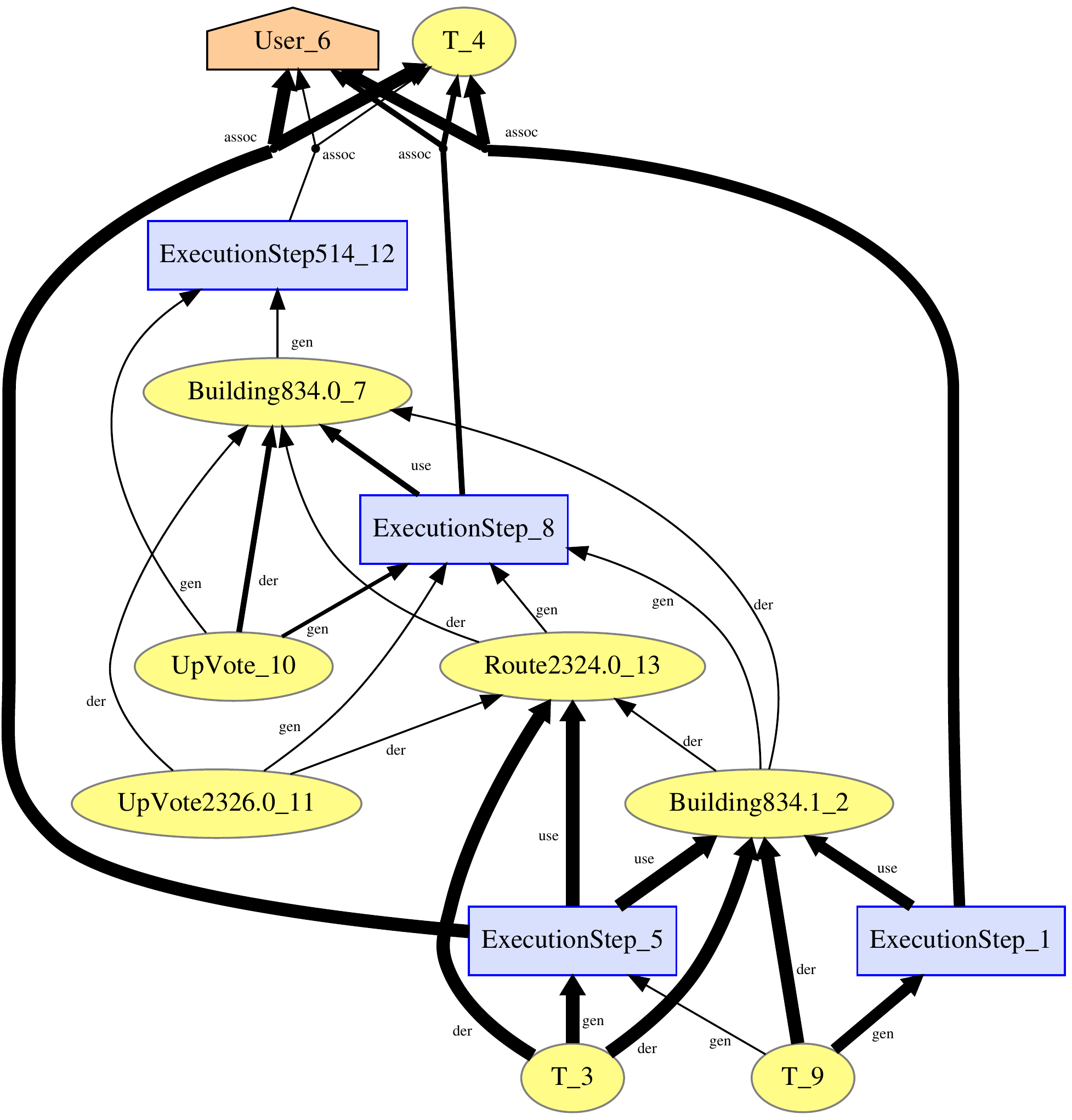}
\end{center}
\caption{The scatter plot on the left-hand side displays the various
  types against the number of nodes for each types. The provenance
  summary on the right-hand side is obtained by application of
  $\APT(1)$ to \col. The thicker the edge in the summary, the more edge instances
  between nodes with corresponding types. 
Left and right sides are available online 
\href{https://eprints.soton.ac.uk/364726/16/scatter.pdf}{here}
and
\href{https://eprints.soton.ac.uk/364726/17/collabmap-visual-1.pdf}{here}, respectively.
}\label{scatter}
\end{figure}

\begin{figure}[htb]
\begin{center}
\provResource{https://eprints.soton.ac.uk/364726/18/collabmap-visual-1-pattern.pdf}
\provFromData{https://eprints.soton.ac.uk/364726/17/collabmap-visual-1.pdf}
\includegraphics[width=0.25\hsize]{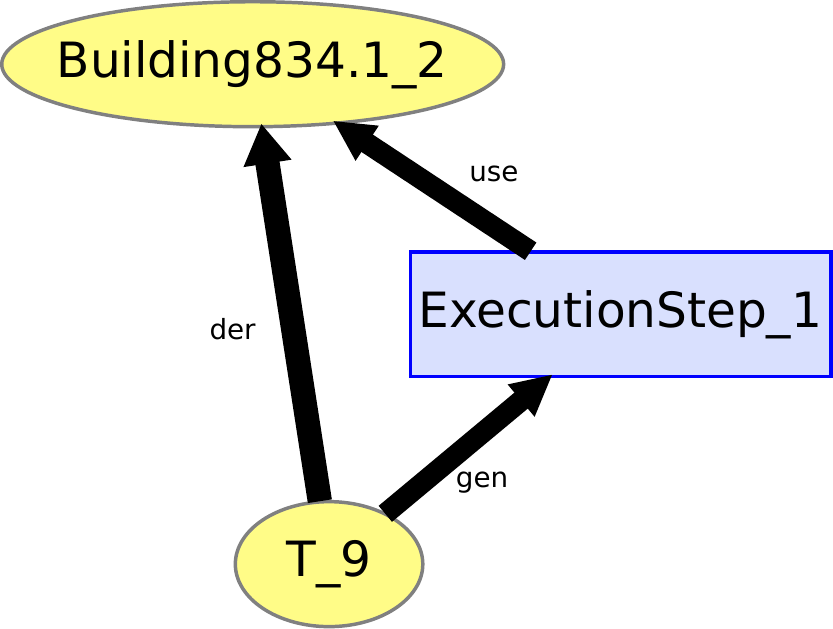}\hspace{2cm}
\provResource{https://eprints.soton.ac.uk/364726/19/collabmap-visual-1-pattern2.pdf}%
\provFromData{https://eprints.soton.ac.uk/364726/17/collabmap-visual-1.pdf}%
\includegraphics[width=0.25\hsize]{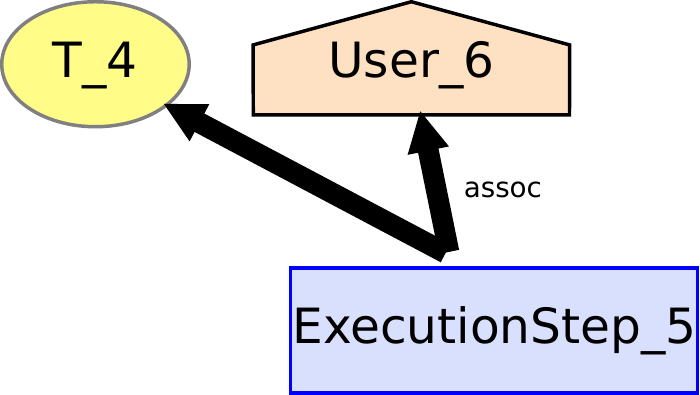}
\end{center}
\caption{Example of  Two Graph Patterns Occurring  6 Times in \col}\label{pattern}
\end{figure}

\smallskip
\Analysis We note that \APT itself does not detect
patterns, but instead \APT helps identify types that occur in repeated
patterns.  Due to space shortage, we focus  on \col only; the analysis could
be performed on other graphs with a similar outcome.
Figure~\ref{scatter} (left)
illustrates a provenance summary produced by $\APT(1)$ on \col, and
the corresponding scatter plot (types vs node occurrences).  
Figure~\ref{scatter} (right) displays a provenance summary, in which the
thickness of edges is proportional to the number of edges in the
original graph instance. Figure~\ref{pattern} displays examples of
repeated patterns. We see that there are 6 and 7 occurrences for types
{\tt ExecutionStep\_1}, {\tt ExecutionStep\_5} and {\tt T\_9}, respectively, in
Figure~\ref{scatter} (left), whereas there are 6 and 10 occurrences
for nodes of type {\tt T\_4} and {\tt User\_6}, respectively.$\Box$

\bigskip

\section{Discussion}\label{discussion:section}

The \APT transformation results in a summary that includes nodes and
edges frequencies. Figure~\ref{scatter} (right) illustrates this
information in a simple graphical way.  We believe that this
information could be the basis of a provenance visualization tool,
but developing such a tool would require substantial work beyond the
scope of this paper.  Instead, we had a discussion with the provenance
expert behind the generation of \col provenance, explaining \APT and
discussing some graphical illustrations, including the one in
Figure~\ref{scatter} (right).  This section summarises some highlights
of the discussion, which was centered around the need to make provenance understandable (see Section~\ref{requirements:section}).

\paragraph{\APT output is helpful to derive a narrative from a provenance graph.\/}
The bold edges are useful to express a narrative. ``A building ({\tt
  Building834.0\_7}) is identified ({\tt ExecutionStep514\_12}), and
voted upon ({\tt ExecutionStep\_8}); a route ({\tt Rou\-te2324.0\_13})
drawn and voted upon ({\tt ExecutionStep\_5}); and, a route set ({\tt
  Buil\-ding834.1\_2}) defined and voted upon ({\tt
  ExecutionStep\_1}).''  A visualization tool leveraging \APT but also
application types would be able to expose such a narrative in a
compelling way.  Thus, we conclude that \APT provides solid
foundations to address the introduction's Question Q1, reflected in Requirement~\ref{essence:requirement}.

\paragraph{\APT helps get a good insight in the way provenance is modelled.\/}
As we were discussing the high-level narrative, our attention was
drawn to an activity type ({\tt ExecutionStep\_8}) that results in a
vote ({\tt UpVote\_10}) and a route ({\tt Route\-2324.0\_13}).  Upon
further investigation, it was discovered that this type encompasses
two different kinds of activities (voting and drawing route,
respectively) of the original provenance graph, but there was nothing
to distinguish those activities since they have the same provenance
type: namely, they they use a building and are associated with some
user, according to some plan.  Given that all the plans are aggregated
as entities that do not have any ancestor, they all appear as a simple
type ({\tt T\_8}).  By distinguishing these types of plans, distinct
activity types would be produced in place of {\tt
  ExecutionStep\_8}. Again, this discussion shows that \APT  helps
 address Requirement~\ref{essence:requirement}.

\paragraph{\APT helps detect outliers.\/}
Large edges in the provenance summary are indicative of repeated
patterns (Hypothesis~\ref{hypo:repeated:patterns}).  Hence, occurrence
of a thin edge within a repeated pattern is highlighting the presence
of an unusual phenomenon. For instance, in Figure~\ref{scatter}
(right), {\tt T\_9} is suspiciously generated from {\tt
  ExecutionStep\_5} with a thin edge. An investigation shows that this
edge links a negative vote on a route set to a voting activity on a
route: when a route is voted negatively, so is the route set it is
contained in. Such phenomenon reveals something special that occurred
at execution time, or alternatively it highlights some specific
provenance modelling by the implementer. Hence, the weights included
in \APT summaries are a good mechanism to address the introduction's
Question Q3 (i.e. Requirement~\ref{outliers:requirement}).

\section{Related Work}\label{related:section}

There are relevant summarization techniques for general graphs, without focus on provenance.
A Data\-Guide~\cite{Goldman:vldb1997} consists of a
dynamically-generated summary of a graph-struc\-tu\-red database.  A
DataGuide for a source object $s$ is an object $d$ such that every
label path of $s$ has exactly one data path instance in $d$
(conciseness), and every label path of $d$ is a label path of $s$
(accuracy). Incremental and non-incremantal algorithms are proposed to
compute DataGuides, and their performance is studied.  Also,
techniques to optimise queries based on DataGuides are investigated.
By construction DataGuides never include information that does not
exist in the data.  On the other hand, \APT aggregate nodes that have
the same provenance types, hereby potentially creating loops in the
output: such loops could correspond to arbitrarily long paths
in the original graph, a fundamental difference between the two approaches. Like \APT, 
the construction of DataGuides has an exponential upper bound for cyclic graphs.

DataGuide's ancestor, Representative Objects (RO)~\cite{ilprints269},
is a form of summary that could be computed for a given graph
database. One of its variants is a $k$-representative, which
limits the summary to path expressions of length $k+1$. While the
definition of RO($k$) is totally different from $\APT(k)$, both have
in common the focus on paths of length $k$.

In subsequent work, Goldman and Widom~\cite{ilprints412} consider
Approximate Data\-Guides, by lifting the accuracy constraint in the
definition, and hereby not requiring every label path of $d$ to be a
label path of $s$.  They use a notion of set similarity
based on the idea that two similar sets have a proportion of common
elements above some threshold. Wang~\etal~\cite{Wang:2000} study a
variant of Approximate DataGuides by maximising a utility function
over a clustering of nodes. Related to our work is their taking into
account of incoming and outgoing edge labels, though they focus on the
size of label sets, rather than the labels themselves.

To optimise queries for the data web, Jarrar and
Dikaiakos~\cite{Jarrar:TKDE12} introduce two notions of graph
signature: the O-Signature (resp. I-Signature) is a summary of a graph
such that nodes that have the same outgoing (resp. incoming) paths are
grouped together. This notion is similar to a
1-Index~\cite{Milo:ICDT99}, a computationally-efficient refinement
over a language equivalence relation.  The key differentiator is that paths of arbitrary length are considered in
\cite{Jarrar:TKDE12,Milo:ICDT99}, whereas \APT limits itself to paths
of length $k$.

Buneman~\etal~\cite{Buneman:1997} define the notion of graph schema
for a graph database. They further introduce the notion of a database
conforming to a schema by a generalization of similarity. Intuitively,
the set of label paths in a schema is a superset of label paths in the
original graph.

Nestorov~\etal~\cite{Nestorov:1998} introduce the notion of
approximate typing, and a measure in terms of defects (number of edges
ignored or to be added to be able to check conformance).  In that
sense, the summary produced by \APT is perfect. They also use a
clustering technique to reduce the number of types, while keeping the
number of defects minimum.

Observing that existing graph summarization methods are mostly
statistical (e.g., degree distribution,
distance, and clustering coefficients~\cite{Newman:2010:NI:1809753}),
Tian~\etal~\cite{Tian:2008} propose two graph summarization techniques
allowing resolutions of summaries to be better controlled.  SNAP
produces a summary graph by grouping nodes based on user-selected node
attributes and relationships.  SNAP produces a
graph partitioning where all nodes in a grouping are homogeneous in
terms of some user-selected attributes and relations; the partitioning
is optimal in the sense that it contains a minimal number of nodes.  A
variant of this operation, $k$-SNAP require most nodes (as opposed to
all) of a grouping to be involved in the selected relation, whereas
they still all have the same attributes; users can select the number
$k$ of groups in the summary. 
\APT proceeds by converting provenance-related relations into
provenance-type attributes.  The grouping produced by \APT is then
equivalent to a SNAP operation over provenance-type attributes.  The
originality of \APT is that it considers relation paths of lengh $k$,
whereas SNAP focuses on direct relations.

Approximations, such as those described
in~\cite{Tian:2008,Nestorov:1998,ilprints412,Wang:2000,Milo:ICDT99}, could be
considered if the number of types produced by \APT is too large. But all
come with non-trivial computational overheads.

In the context of Business Activity Monitoring, process mining
consists of extracting information about processes from transaction
logs.  Transactions logs typically are a strict subset of provenance
information.  The type of processes that can be extracted can be
represented as Petri Nets~\cite{vanDerAalst:TKDE04}.

Semantic substrates is a technique to group nodes into rectangular
regions, and lay them within each region according to user-chose
attributes. Node aggregation~\cite{16937} is used to replace all the
nodes in a grid cell with a single metanode. Likewise,
PivotGraph~\cite{Wattenberg:2006} is a technique to visualize graph
according to attributes selected by users.  The
techniques~\cite{16937,Wattenberg:2006} are complementary with \APT:
\APT exposes provenance types as attributes, and these techniques
empower the user to select which attributes to render, and offer
original layouts.   Koop~\etal~\cite{Koop:Vis13} propose a
method to summarise graph collections: they use domain-specific
comparison functions to collapse similar nodes and edges, with the aim
§of producing more compact representations of such collections.
An interesting study would be to focus on the 
suitability of all these visualization methods for provenance.

Means of abstracting provenance traces have been considered. ``User
views'', defined as a partition of tasks in
a workflow specification~\cite{Cohen-Boulakia:CCPE08}, provide the means to selectively identify
what aspect of a provenance trace should be exposed to users. This
approach differs from our work since our summaries are construted
automatically without requiring an explicit workflow. Likewise,
``abstract provenance graph'' \cite{Zinn:IPAW2010} are derived by
static analysis of workflows.  In addition, techniques are proposed to
further summarise provenance graphs, based on time information and the
structure of the worklow.  Provenance graph abstraction by node
grouping~\cite{Missier:TR2013} is a technique by which a set of nodes
in a \PROV-compliant provenance graph is replaced by a new abstract
node with a view to obfuscate provenance;  privacy policies are
used to identify the nodes to group.




\section{Concluding Remarks and Future Work}\label{conclusion:section}

In this paper, we have presented a summarisation technique for
provenance graphs.  The approach consists of converting provenance
paths, up to some length $k$ to node attributes (referred to as
provenance types), and grouping nodes that have the same provenance
types.  The summary also contains numeric values representing the
frequency of nodes and edges in the original graph.

A complexity analysis of the algorithm shows that it is linear in the
size of the graph, and potentially exponential in the maximum path
length $k$. Such a type of complexity is typical of related work. The
positive side of our work is that our quantitative evaluation shows
that the algorithm is perfectly tractable since useful summary graphs
can be obtained with small values of $k$, and it was shown that for
$k$ greater than the Maximum Finite Distance, the size of the summary
saturates. 

We also introduced a notion of conformance to a summary, which
captures the idea that the summary includes all possible traversals
that can be performed in a graph.  We have illustrated how such
conformance can be implemented by means of an OWL2 consistency check.

A qualitative discussion of the approach, based
on a sample visualization based on the summarization, has shown
that the approach has a good potential to detect anomalies and
outliers.  Overall, we have demonstrated that \APT suitably addresses
three use cases, formulated as questions in the introduction of the
paper and expressed as requirements for summarisation.

With this paper, we have opened up a whole area of research in
summarization techniques for provenance graphs, and their application
to conformance checking and visualization. Our future work will seek
to develop efficient algorithms for conformance checking. In addition,
we will seek to investigate the incremental aspect of the approach:
being able to adjust summaries and being able to check conformance
incrementally as provenance graphs are extended.  Furthemore, an issue
that is particularly related is to find mechanisms to help users
identify base types to construct provenance types.  Finally, while the
approach is developed in the context of provenance, it could have
potential applications for any form of graphs; future research would
have to clarify the intuition associated with so-called provenance
types.

\paragraph{Acknowledgements}
This work is funded in part by the EPSRC \provProject{SOCIAM (EP/J017728/1)}{http://www.sociam.org/}{http://www.epsrc.ac.uk/} and
\provProject{ORCHID (EP/I011587/1)}{http://www.orchid.ac.uk/}{http://www.epsrc.ac.uk/} projects, 
the \provProject{FP7 SmartSociety (600854)}{http://www.smart-society-project.eu/}{http://cordis.europa.eu/fp7/} project,
and the ESRC 
\provProject{eBook (ES/K007\-246/1)}{http://www.bristol.ac.uk/cmm/research/ebooks/}{http://www.esrc.ac.uk/} project.  Thanks to Trung Dong Huynh
for feedback on a draft of this paper.

\provBibliography

\bibliography{gam15} 
\bibliographystyle{eptcs}

\renewcommand{\datastatement}{Provenance datasets referred to by this paper are listed below.}

\provDatagraphy

\datagraphy{data}
\datagraphystyle{eptcs-data}

\provEmbed

\end{document}